\documentclass[aps,reprint,groupedaddress,showpacs]{revtex4-1}

\usepackage{euscript,amsmath,amssymb,amsfonts,graphicx,color}
\usepackage{epstopdf}

\newcommand{\e}{{\rm e}}
\renewcommand{\d}{{\rm d}}
\newcommand{\pd}{\partial}

\newcommand{\mc}{\mathcal }
\newcommand{\ve}{\varepsilon}

\newcommand{\bd}{{\mathbf \Delta}}

\begin{document}

\title{Stochastic synchronization of neural activity waves }
\author{Zachary P. Kilpatrick}
\email{zpkilpat@math.uh.edu}
\affiliation{Department of Mathematics, University of Houston, Houston, TX 77204}
\date{\today}

\begin{abstract}

We demonstrate that waves in distinct layers of a neuronal network can become phase-locked by common spatiotemporal noise. This phenomenon is studied for stationary bumps, traveling waves, and breathers. A weak noise expansion is used to derive an effective equation for the position of the wave in each layer, yielding a stochastic differential equation with multiplicative noise. Stability of the synchronous state is characterized by a Lyapunov exponent, which we can compute analytically from the reduced system. Our results extend previous work on limit-cycle oscillators, showing common noise can synchronize waves in a broad class of models.

\end{abstract}

\maketitle

\noindent
{\bf Introduction.} Nonlinear waves arise in many physical, biological, and chemical systems including non-equilibrium reactions \cite{jakubith90}, shallow water \cite{huang99}, bacterial populations \cite{danino10}, epidemics \cite{cummings04}, and cortical tissue \cite{wang10}. Phase synchronization of multiple waves can occur when their dynamics are coupled. For example, spiral waves in chemical systems become entrained when coupled diffusively through a membrane \cite{winston91}. Experiments on amoeba populations have also demonstrated entrainment via interactions between wave-emitting centers and spiral waves  of cell density \cite{lee96}.  Common forcing can also synchronize waves at distinct spatial locations. Spatiotemporal analysis of epidemics reveals that both seasonality and vaccination schedule can entrain the nucleation of outbreak waves across geographical space \cite{grenfell01}.  Furthermore, activity recordings from primary visual cortex show that triggering switches during binocular rivalry leads to synchronized wave initiation \cite{lee05}. In total, experimental studies demonstrate a wide array of mechanisms for synchronizing the onset and propagation of waves.

Our goal in this Rapid Communication is to show that stochastic forcing can also entrain the phases of distinct waves. We focus on neural activity waves that arise due to distance-dependent synaptic interactions \cite{ermentrout01,wang10}. Waves of neural activity underlie sensory processing \cite{xu07}, motor action \cite{rubino06}, and sleep states \cite{massimini04}. Proposed computational roles of neural activity waves include heightening the responsiveness of specific portions of a network and labeling incoming signals with a distinct phase \cite{ermentrout01}. Thus, it may be advantageous for waves to be coordinated across multiple brain areas, and we propose that correlated fluctuations may underlie such coordination. Many theoretical and experimental studies have identified ways noise correlations can degrade neural information encoding \cite{cohen11}, but recent work has shown correlated noise can reliably synchronize activity across populations of neurons \cite{ermentrout08}. 

Our analysis extends previous work which showed common noise can synchronize the phases of limit cycle oscillators \cite{pikovsky01,teramae04}. A key observation of these studies is that the synchronous state, where all oscillators have the same phase, is absorbing when each oscillator receives identical noise. Stability of the phase-locked state can then be determined by computing an associated Lyapunov exponent, which is negative for nontrivial phase response curves \cite{teramae04}. As we show in this Rapid Communication, these principles can be applied to waves forced by common spatiotemporal noise. Waves are driven to an attracting synchronous state, where the waves' phases are identical.

Synchronization of neural activity waves across distinct network locations can be important in several behavioral and sensory contexts. For instance, waves of activity in the visual cortex may function like a ``bar code scanner," ensuring a portion of the network is always maximally sensitive to external inputs \cite{ermentrout01}. Since locations in visual space are represented by multiple layers of a network, coordinating background waves across layers could ensure the network is always sensitive at the same visual location in each layer. Thus, our findings implicate an important potential role for large-scale correlations in nervous system fluctuations, which are often deemed a nuisance to cognitive performance \cite{cohen11}.

In this Rapid Communication, we analyze the stochastic dynamics of waves in a pair of uncoupled neural field models driven by common noise. Neural fields are nonlinear integrodifferential equations whose integral term describes the connectivity of a neuronal network \cite{amari77,bressloff12}. Recent studies have considered stochastic versions of neural field equations, formulating the dynamics as Langevin equations with spatiotemporal noise \cite{bressloff12,hutt07}:
\begin{align}
\d u_j(x,t) = \left[ - u_j(x,t) + w*f(u_j) \right] \d t + \ve \d W(x,t), \label{layersys}
\end{align}
where $u_j(x,t)$ is neural activity of population $j=1,2$ at $x \in [ - \pi, \pi]$ at time $t$, synaptic connectivity is described by the convolution $w*f(u)=\int_{-\pi}^{\pi} w(x-y)f(u(y,t)) \d y$, and $f(u)$ is a nonlinearity describing the fraction of active neurons.
Small amplitude ($\ve \ll 1$) spatiotemporal noise $\d W (x,t)$ is white in time and correlated in space, so $\langle \d W (x,t) \rangle = 0$ and $\langle \d W(x,t) \d W(y,s) \rangle = 2 C(x-y) \delta (t-s) \d t \d s$. As noise correlations must thus be even and $2 \pi$-periodic, we write $C(x) = \sum_{k=0}^{\infty} a_k \cos (kx)$. \\
\vspace{-2.5mm}


\begin{figure}
\begin{center} \includegraphics[width=9cm]{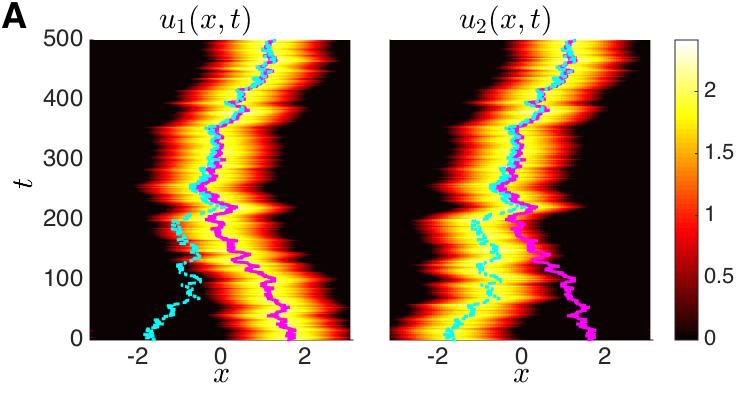} \\ 
\includegraphics[width=4.4cm]{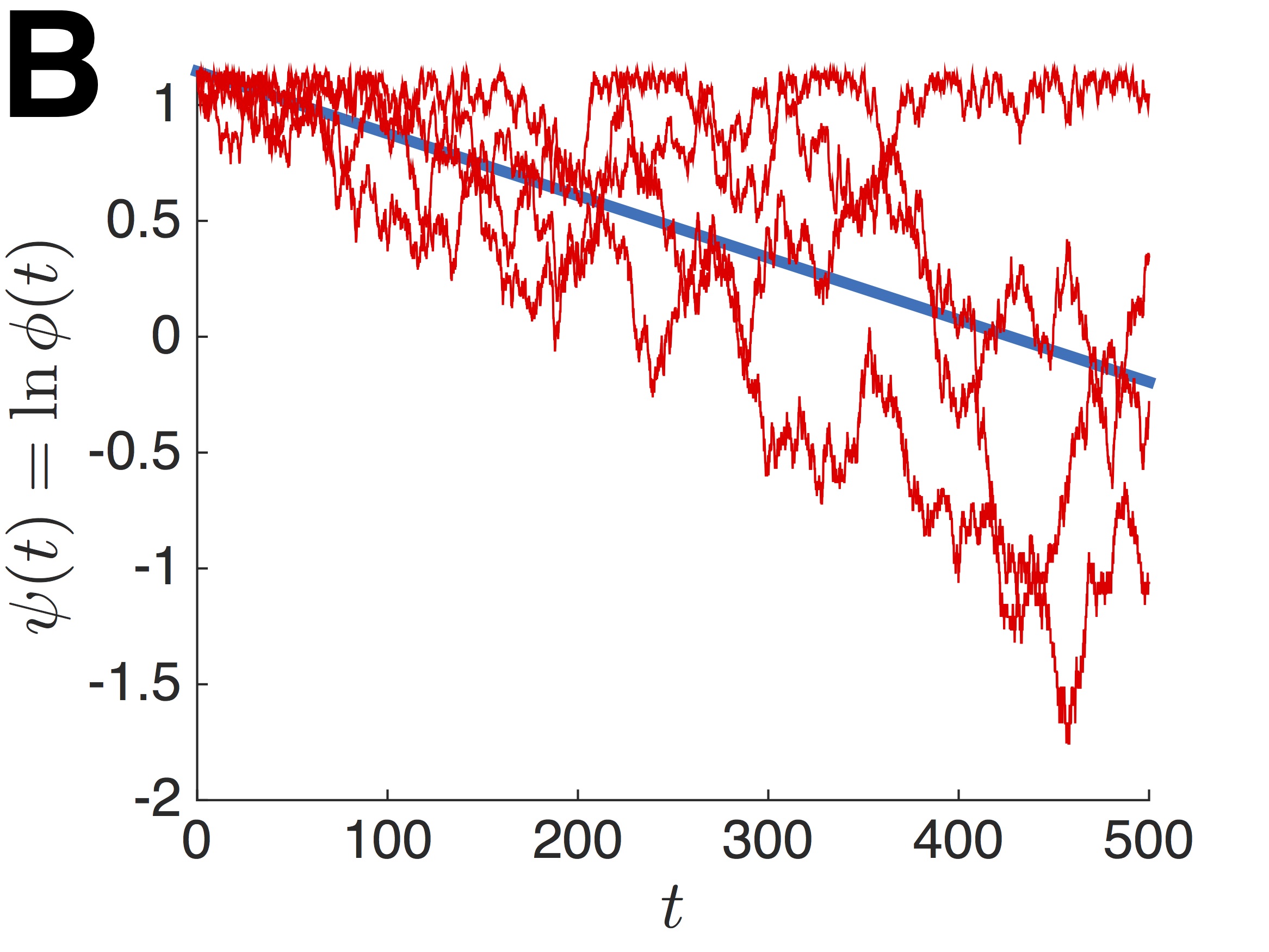} \hspace{-3mm}\includegraphics[width=4.4cm]{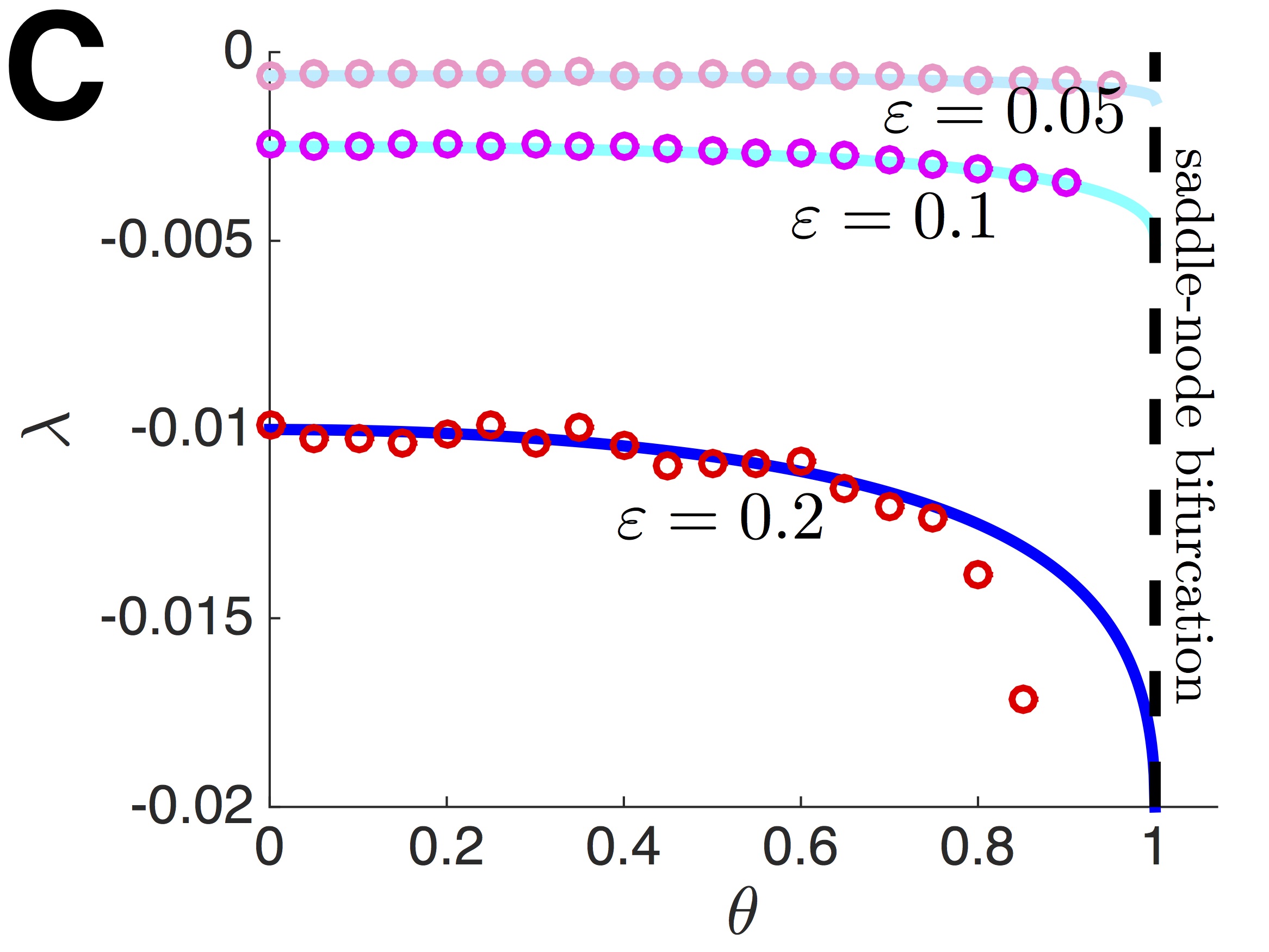} \end{center}
\vspace{-6mm}
\caption{(color online) Common noise-induced phase-locking of bumps evolving in two uncoupled stochastic neural fields, Eq.~(\ref{layersys}) with $w(x)=\cos(x)$, and $C(x)=\cos(x)$. ({\bf A}) In a single realization, noise eventually drives bump positions $\Delta_1(t)$ (solid line) and $\Delta_2(t)$ (dashed line) to the absorbing state $\Delta_1(t)=\Delta_2(t)$. Here, $f(u)=H(u-0.5)$ and $\ve = 0.01$. ({\bf B}) Realizations of the log phase-difference $\ln |\phi| = \ln |\Delta_1-\Delta_2|$ (thin lines) compared to theory $\ln |\phi (t)| = \ln |\phi (0)| +  \lambda t$ (thick line) given by the formula Eq.~(\ref{lyapefin}). ({\bf C}) Numerically calculated Lyapunov exponents $\lambda$ (circles) are well approximated by Eq.~(\ref{lyapefin}) (line), increasing in amplitude $|\lambda|$ as $\theta$ moves toward the saddle-node bifurcation of bumps \cite{numerics}. }
\vspace{-4mm}
\label{synchbump}
\end{figure}

\noindent
{\bf  Stationary bumps.} We begin by demonstrating noise-induced wave synchronization  for a single realization of the system Eq.~(\ref{layersys}) in Fig. \ref{synchbump}{\bf A} \cite{numerics}. Here $w(x-y) = \cos (x-y)$ is an even symmetric lateral inhibitory weight function, known to lead to stable stationary bump solutions in the unperturbed system ($\ve \equiv 0$) \cite{amari77}. Additive spatiotemporal noise causes each bump to wander diffusively about the spatial domain. Once both waves' positions $\Delta_1$ and $\Delta_2$ meet, they are phase-locked for the remainder of the simulation, since both layers receive identical noise and obey the identical governing Eq.~(\ref{layersys}).

To analyze this behavior, we first note that the unperturbed system, $\ve \to 0$ in Eq.~(\ref{layersys}), has stationary bump solutions $u_j(x,t) = U(x - x_j)$ ($j=1,2$) satisfying $U(x) = \int_{ - \pi}^{\pi} w(x- y) f(U(y)) \d y$ \cite{ermentrout98,kilpatrick13}. There is a degeneracy in the position of each bump's center of mass $x_j$, since the waves are neutrally stable to translating perturbations \cite{kilpatrick13}, and noise will cause waves to stochastically wander about their mean position \cite{panja04,bressloff12b}.

We now derive effective equations for the bumps' response to noise ($\ve >0$) by utilizing a perturbation expansion that tracks the stochastically varying position $\Delta_j (t)$ of each bump and fluctuations in the wave profiles $\Phi_j (x, t)$, so $u_j(x,t) = U(x - \Delta_j(t)) + \ve \Phi_j ( x - \Delta_j(t) , t) + {\mc O}( \ve ^2 )$. Bumps in each layer may begin at distinct locations $\Delta_j(0) = x_j$, but we will show that when receiving common noise, their locations become synchronized  in the long time limit $\lim_{t \to \infty} |\Delta_1(t) - \Delta_2(t)| = 0$.

Plugging this ansatz into Eq.~(\ref{layersys}) and expanding to ${\mc O}( \ve)$, we have the linear stochastic equations
\begin{align}
\d \Phi_j - {\mc L} \Phi_j \d t = \ve^{-1} \d \Delta_j U'  + \d W ( x + \Delta_j, t), \label{phiexp}
\end{align}
for $j=1,2$, where ${\mc L} u = - u + w*\left[ f'(U) \cdot u \right]$ is a linear operator defined for integrable functions $u(x)$. Both layers $j=1,2$ receiving identical noise, so we must keep track of each wave's relative phase $\Delta_j$. A solution to Eq.~(\ref{phiexp}) exists if we require the right hand side be orthogonal to the nullspace of the adjoint operator ${\mc L}^* p = - p + f'(U)  \cdot w*p$. Define $V$ to be a one-dimensional basis of ${\mc N}({\mc L}^*)$. Taking the inner product of Eq.~(\ref{phiexp}) with $V$, we find
\begin{align}
\langle V (x) , \ve^{-1} \d \Delta_j U_j' (x)  + \d W (x + \Delta_j, t) \rangle = 0,  \label{solvcond}
\end{align}
so $\Delta_j$ obeys a Langevin equation with multiplicative noise
\begin{align}
\d \Delta_j = \ve \frac{ \int_{- \pi}^{\pi} V (x) \d W (x+ \Delta_j, t) \d x}{\int_{- \pi}^{\pi} V(x) U'(x) \d x },  \ \ \ j=1,2.  \label{langeq}
\end{align}
We can represent $W( x, t)$ using the Fourier expansion
\begin{align}
W(x,t) = a_0 X_0 + \sum_{k=1}^{\infty} a_k \left[ X_k \cos (kx) + Y_k \sin (kx) \right],  \label{wnosfour}
\end{align}
where $X_k$ and $Y_k$ are white noise processes $\langle X_k(t) \rangle = \langle Y_k(t) \rangle = 0$ and $\langle X_k(t) X_l(t) \rangle = \langle Y_k(t) Y_l(t) \rangle = \delta_{kl} \delta (t-s)$; $\delta_{kl}$ is the Kronecker delta function. Using trigonometric identities, we can express
\begin{align}
\d \Delta_j = \sqrt{2} \ve  \d {\mc W}(\Delta_j,t), \ \ \ j=1,2,  \label{ddelfour}
\end{align}
where ${\mc W}(\Delta, t) = \sum_{k=1}^{\infty} \left[ b_{+k} \cos (k \Delta) X_k + b_{-k} \sin (k \Delta) Y_k \right]$, and the $X_0$ vanishes since $\int_{- \pi}^{\pi} V(x) \d x \equiv 0$ and
\begin{align*}
b_{\pm k} &= \frac{a_k \int_{- \pi}^{\pi} V(x) \left[ \cos ( k x ) \pm \sin (kx) \right] \d x}{\int_{- \pi}^{\pi} V (x) U'(x) \d x}.
\end{align*}
Thus, we have reduced Eq.~(\ref{layersys}) to a system describing two phase oscillators perturbed by weak noise, Eq.~(\ref{ddelfour}). The coefficients $b_{\pm k}$ describe the relative contributions of each term of the Fourier series, Eq.~(\ref{wnosfour}), to the phase-dependent sensitivity of the oscillators to noise.

The synchronized solution $\Delta_1(t) = \Delta_2(t)$ to Eq.~(\ref{ddelfour}) is absorbing since the the right-hand sides of both equations ($j=1,2$) will subsequently be identical. To assess stability of the absorbing state, we compute the associated Lyapunov exponent $\lambda$. Proper calculation requires translating Eq.~(\ref{ddelfour}) into its equivalent Ito formulation \cite{gardiner04}
\begin{align}
\d \Delta_j= \ve^2 R( \Delta_j) \d t +  \sqrt{2}   \ve \d {\mc W}(\Delta_j,t), \ \ \ j=1,2,  \label{ddelito}
\end{align} 
where the Ito Eq.~(\ref{ddelito}) introduces the drift term
\begin{align}
R( \Delta) =  \sum_{k=1}^{\infty} \left[ b_{-k}^2  - b_{+k}^2 \right] k \sin (k \Delta) \cos (k \Delta),  \label{itodrift}
\end{align}
accounting for the fact that the correlation between state variables and noise terms, present in the Stratonovich Eq.~(\ref{ddelfour}), subsequently vanishes. We proceed by formulating the variational equation for the perturbative phase difference $\phi (t) = \Delta_1 (t) - \Delta_2(t)$ ($|\phi| \ll 1$), which can be derived from Eq.~(\ref{ddelito}), so
\begin{align}
\d \phi  = \ve^2 R'(\Delta) \phi \d t + \sqrt{2} \ve  \phi \d {\mc Y}(\Delta , t),  \label{pdiffeq}
\end{align}
where ${\mc Y}(\Delta,t) =  \sum_{k=1}^{\infty} k \left[ b_{-k} \cos (k \Delta) X_k - b_{+k} \sin (k \Delta) Y_k \right]$ and $\Delta$ obeys Eq.~(\ref{ddelito}). Defining $\psi  = \ln \phi$ and appealing to Ito's formula, we can rewrite Eq.~(\ref{pdiffeq}) as
\begin{align}
\d \psi  = \ve^2 \left[ R'(\Delta) - S ( \Delta) \right] \d t + \sqrt{2} \ve  \d {\mc Y}(\Delta , t), \label{psidife}
\end{align}
where
\begin{align}
S( \Delta )  = \sum_{k=1}^{\infty} k^2 \left[ b_{+k}^2 \sin^2 (k \Delta)  + b_{-k}^2 \cos^2( k \Delta) \right].  \label{sdelta}
\end{align}
Subsequently, we can integrate Eq.~(\ref{psidife}) to determine the mean drift of $\psi (t)$
\begin{align}
\lambda := \ve^2 \lim_{t \to \infty} \int_0^t \left[ R'(\Delta(s)) - S(\Delta(s)) \right] \d s,  \label{lyape1}
\end{align}
which is also the mean rate of growth of $\phi (t)$. The phase difference $\phi (t)$ will tend to decay (grow) if $\lambda <0$ ($\lambda>0$) and synchrony will be stable (unstable). Utilizing ergodicity of Eq.~(\ref{psidife}), we can equivalently compute $\lambda$ with the ensemble average across realizations of ${\mc Y}(\Delta, t)$, so \cite{teramae04}
\begin{align}
\lambda = \ve^2 \int_{ - \pi}^{\pi} P_s( \Delta ) \left[ R'(\Delta) - S(\Delta) \right] \d \Delta,  \label{lyape2}
\end{align}
where $P_s(\Delta)$ is the steady state distribution of $\Delta$. Since noise is weak ($\ve \ll 1$), we can approximate the distribution as constant, $P_s(\Delta) = 1/(2 \pi)$. Applying this to Eq.~(\ref{lyape2}), we find the first term of the integrand vanishes since $R( \pi ) = R(- \pi)$ according to Eq.~(\ref{itodrift}), so the Lyapunov exponent is approximated by the formula
\begin{align}
\lambda &= - \frac{\ve^2}{2 \pi} \int_{ - \pi}^{\pi} S ( \Delta ) \d \Delta = - \frac{\ve^2}{2} \sum_{k=1}^{\infty} k^2 \left[ b_{+k}^2 + b_{-k}^2 \right].   \label{lyapefin}
\end{align}
Note that as long as $b_{+k} \neq 0$ or $b_{-k} \neq 0$ for some $k< \infty$, we expect $\lambda <0$, so the phase-locked state $\Delta_1(t) = \Delta_2(t)$ will be linearly stable.

We now compare the analytical result Eq.~(\ref{lyapefin}) to results from numerical simulations of Eq.~(\ref{layersys}). Explicit calculations are straightforward in the case of a Heaviside nonlinearity $f(u) = H(u - \theta)$; cosine weight function $w(x) = \cos (x)$; and cosine spatial noise correlations $C(x) = \cos (x)$. Stable stationary bump solutions are given by the formulas $U(x) = 2 \sin a \cos (x)$ and $U(\pm a) = \theta$, and the null vector $V (x) = \delta (x-a) - \delta(x+a)$ \cite{kilpatrick13}. Coefficients of the noise Fourier components in Eq.~(\ref{wnosfour}) are $a_k =\delta_{k1}$, so $b_{\pm 1} = \mp 1/[\sqrt{1+ \theta} + \sqrt{1- \theta}]$ and $b_{\pm k} \equiv 0$, $k \neq 1$. Finally, utilizing Eq.~(\ref{sdelta}) along with Eq.~(\ref{lyapefin}), we have
\begin{align}
\lambda = -\frac{\ve^2}{2} \left[ b_{+1}^2 + b_{-1}^2 \right] = - \frac{\ve^2}{2+ 2\sqrt{1- \theta^2}}.  \label{leHbump}
\end{align}
Results from numerical simulations in Fig. \ref{synchbump}{\bf B},{\bf C} corroborate with Eq.~(\ref{leHbump}), showing the Lyapunov exponent's magnitude $|\lambda|$ increases with noise intensity $\ve^2$ and threshold $\theta$. We note that our theoretical approximation breaks down as $\theta$ increases and the system nears a saddle-node bifurcation at which the stable/unstable branch of bump solutions annihilate in the noise-free system \cite{amari77,ermentrout98,kilpatrick13}. Furthermore, the amplitude $|\lambda|$ increases as the parameter $\theta$ is increased towards this bifurcation.

\begin{figure}
\begin{center} \includegraphics[width=3.55cm]{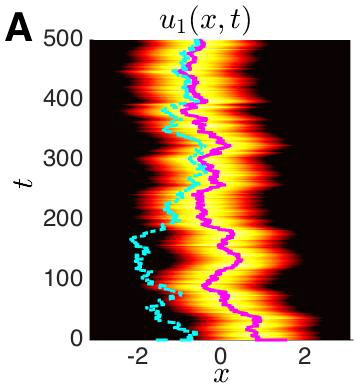} \hspace{3mm}\includegraphics[width=4.65cm]{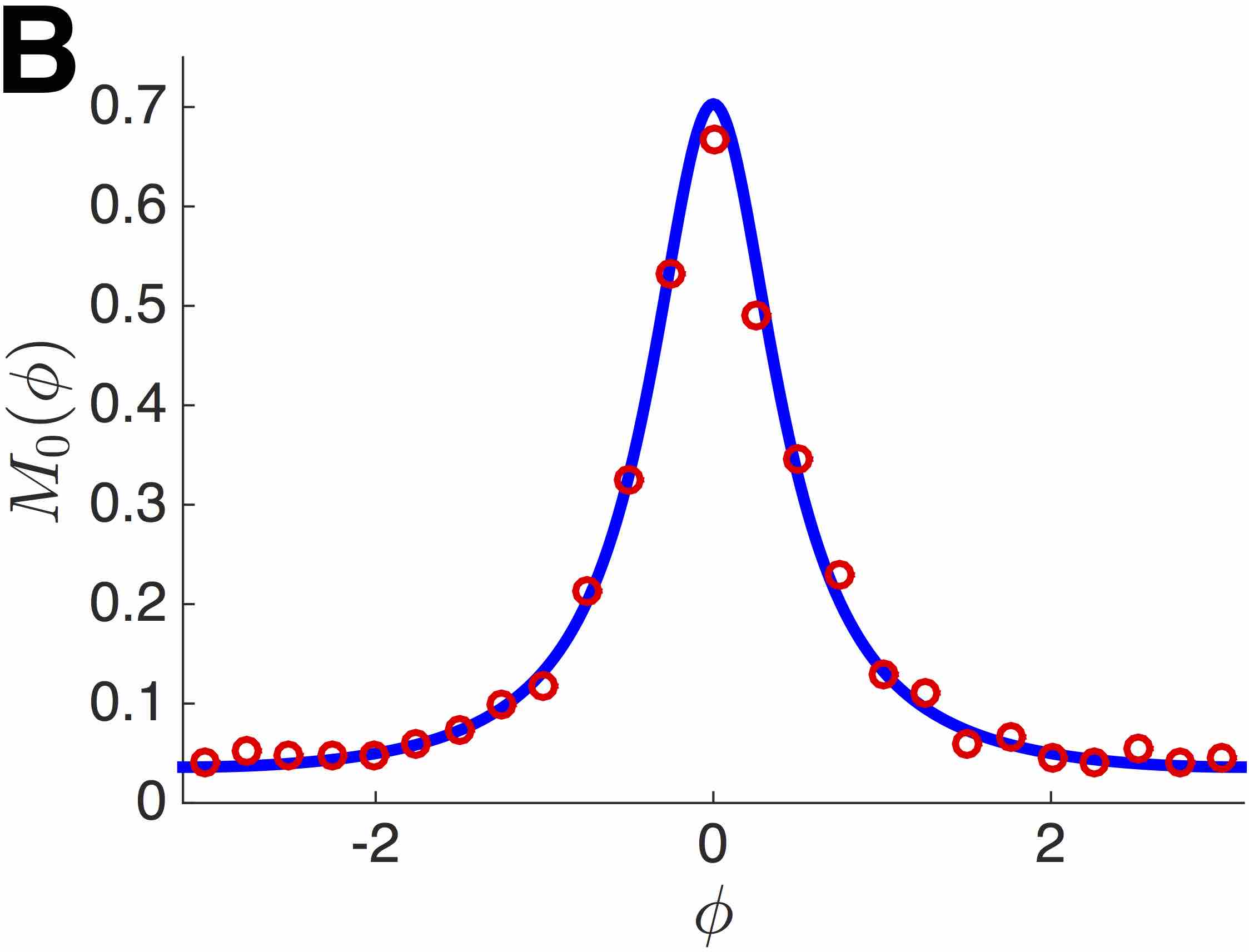}\end{center}
\vspace{-6mm}
\caption{(color online) Independent noise prevents complete phase-locking (see Appendix). ({\bf A}) Noise correlations drive bumps close to one another ($\Delta_1(t) \approx \Delta_2(t)$), but independent noise prevents their remaining in the phase-locked state. ({\bf B}) Stationary density $M_0(\phi)$ widens for $\chi < 1$ in Eq.~(\ref{M0}) theory (line) and numerics (circles). Parameters are $\theta = 0.5$, $\ve = 0.01$, $\chi = 0.95$, and $C_j(x)=\cos(x)$ ($j=1,2,c$).}
\vspace{-4mm}
\label{indnoise}
\end{figure}

We show the robustness of these results by studying the impact of independent noise (Fig. \ref{indnoise}{\bf A}). To do so, we consider a modified version of Eq.~(\ref{layersys}), $\d u_j = \left[ - u_j + w*f(u_j) \right] \d t + \ve \d \widetilde{W}_j$, where $\widetilde{W}_j$ has an independent component in each layer $j=1,2$ (see Appendix). Extending analysis of limit cycle oscillators \cite{nakao07}, we derive an expression for the stationary density
\begin{align}
M_0(\phi) = \frac{m_0}{g(0) - \chi^2 g(\phi)}  \label{M0}
\end{align}
of the phase difference $\phi = \Delta_1 - \Delta_2$. Here, $\chi \in [0,1]$ is the degree of noise correlation, $g( \Delta) = \sum_{k=1}^{\infty} b_k^2 \cos (k \Delta )$ ($b_k = |b_{\pm k}|$), and $m_0$ is a normalization constant. As $\chi$ is decreased from unity, the stationary density widens, representing the effects of independent noise in each layer. However, the density will still tend to be peaked at $\phi = 0$ (see Appendix). Indeed, our theory, Eq.~(\ref{M0}), is corroborated by numerical simulations (Fig. \ref{indnoise}{\bf B}). \\
\vspace{-2.5mm}

\begin{figure}
\begin{center} \includegraphics[width=3.55cm]{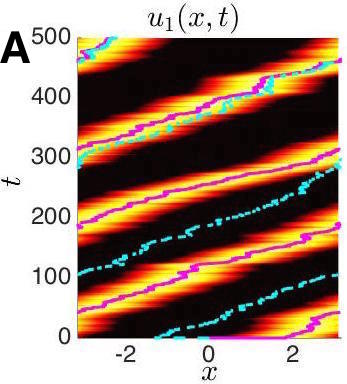} \hspace{-1mm}\includegraphics[width=5cm]{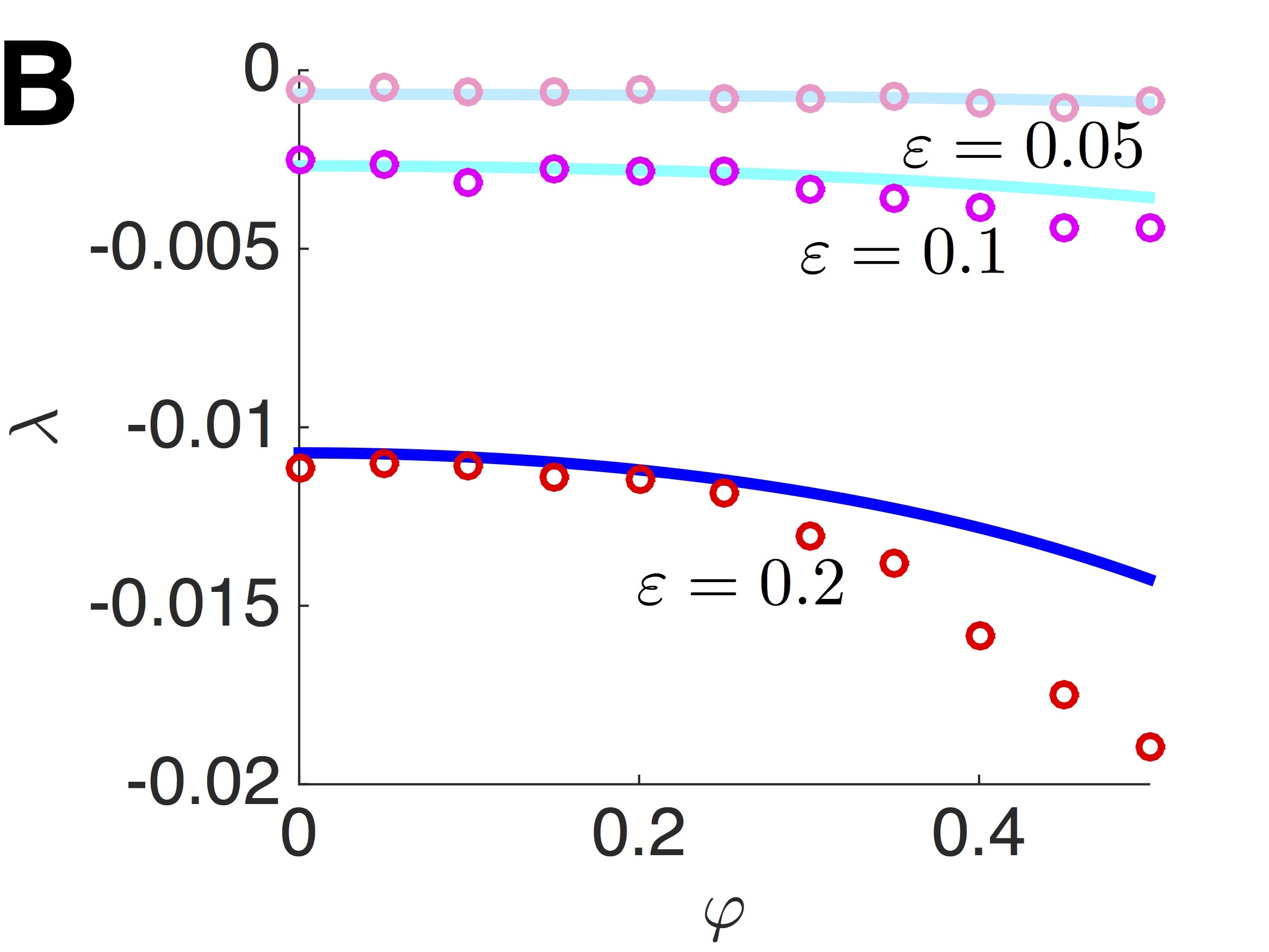}\end{center}
\vspace{-6mm}
\caption{(color online) Noise-induced phase-locking of traveling waves evolving in Eq.~(\ref{layersys}) with $w(x) = \cos (x-\varphi)$. ({\bf A}) Noise perturbations drive wave phases to the phase-locked state $\Delta_1(t)=\Delta_2(t)$. Here $\varphi = 0.2$, while other parameters are as in Fig. \ref{synchbump}.  ({\bf B}) The Lyapunov exponent $\lambda$ calculated from numerical simulations (circles) is approximated by Eq.~(\ref{lyapewaves}) (line), increasing in amplitude $|\lambda|$ with the skewness $\varphi$ of the weight function \cite{numerics}.}
\vspace{-5mm}
\label{synchwaves}
\end{figure}

\noindent
{\bf  Traveling waves.} Our results for stationary bumps can be extended to address stochastic synchronization of traveling waves in networks with asymmetric weights ($w(x) \not\equiv w(-x)$), as in Fig. \ref{synchwaves}{\bf A}. Thus, the unperturbed system, $\ve \to 0$ in Eq.~(\ref{layersys}), will have traveling wave solutions $u_j(x,t) = U(\xi - x_j)$, $\xi = x-ct$, with wave speed $c$ ($j=1,2$), so $- c U'(\xi) = - U(\xi) + w*f(U)$ \cite{ermentrout98,kilpatrick12}. Furthermore, waves will be neutrally stable to translating perturbations, so spatiotemporal noise will cause an effective diffusion of their phases. For $\ve >0$, we apply the ansatz $u_j(x,t) = U(\xi - \Delta_j(t)) + \ve \Phi_j(\xi - \Delta_j(t),t) + {\mc O}(\ve^2)$ with $\Delta_j(t) = x_j$ and will show $\lim_{t \to \infty} |\Delta_1(t) - \Delta_2(t) | = 0$. At ${\mc O}(\ve)$, we find Eq.~(\ref{phiexp}) with corresponding linear operator ${\mc L}u = c u' - u + w*\left[ f'(U) \cdot u \right]$. Solvability is enforced by ensuring the right hand side of Eq.~(\ref{phiexp}) is orthogonal to the nullspace $V$ of the adjoint ${\mc L}^* p = -cp' - p + f'(U) \cdot w(-\xi)*p(\xi)$, yielding the Langevin equation, Eq.~(\ref{langeq}). The Lyapunov exponent $\lambda$ associated with the stability of the absorbing state $\Delta_1(t) = \Delta_2(t)$ is then approximated by Eq.~(\ref{lyapefin}).

To compare our analytical results for traveling waves with numerical simulations, we compute $\lambda$ from Eq.~(\ref{lyapefin}) when $f(u) = H(u-\theta)$, $w(x) = \cos (x - \varphi)$, and $C(x) = \cos (x)$. Stable traveling waves have profile $U(\xi) = \cos \varphi \left[ \sin \xi - \sin ( \xi + a) \right]$, width $a = \pi - \sin^{-1} \left[ \theta \sec \varphi \right]$ defined by thresholds $U(\xi_1) = U(\xi_2) = \theta$ where $\xi_1=\pi-a$ and $\xi_2 = -\pi$, and speed $c = \tan \varphi$ \cite{kilpatrick12}. The null vector can also be computed explicitly:
\begin{align*}
V( \xi ) =& \sum_{k=1}^2 (-1)^k \left[ H( \xi - \xi_k ) + \frac{\coth (\pi/c) - 1}{2} \right] \e^{(\xi_k - \xi)/c}.
\end{align*}
Fourier coefficients of ${\mc W}(\Delta,t)$ in Eq.~(\ref{ddelfour}) are thus given $b_{\pm 1} = (1 \mp c)/2 - (c \pm 1) \sin a/[2 \cdot (1- \cos a)]$ and $b_{\pm k} \equiv 0$, $k \neq 1$, so we compute Eq.~(\ref{lyapefin}), finding
\begin{align}
\lambda = - \frac{\ve^2}{2} \left[ b_{+1}^2 + b_{-1}^2 \right] = -\frac{\ve^2 (1+\cos a)}{2 \theta^2},  \label{lyapewaves}
\end{align}
comparing with numerical results in Fig. \ref{synchwaves}{\bf B}. \\
\vspace{-2.5mm}

\begin{figure}
\begin{center} \includegraphics[width=3.5cm]{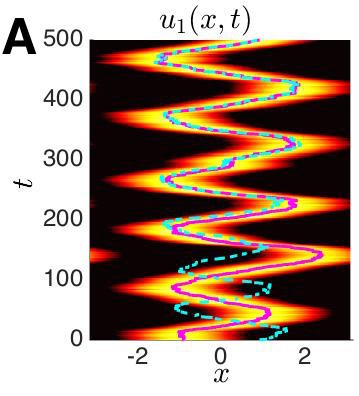}  \hspace{-1mm}\includegraphics[width=5cm]{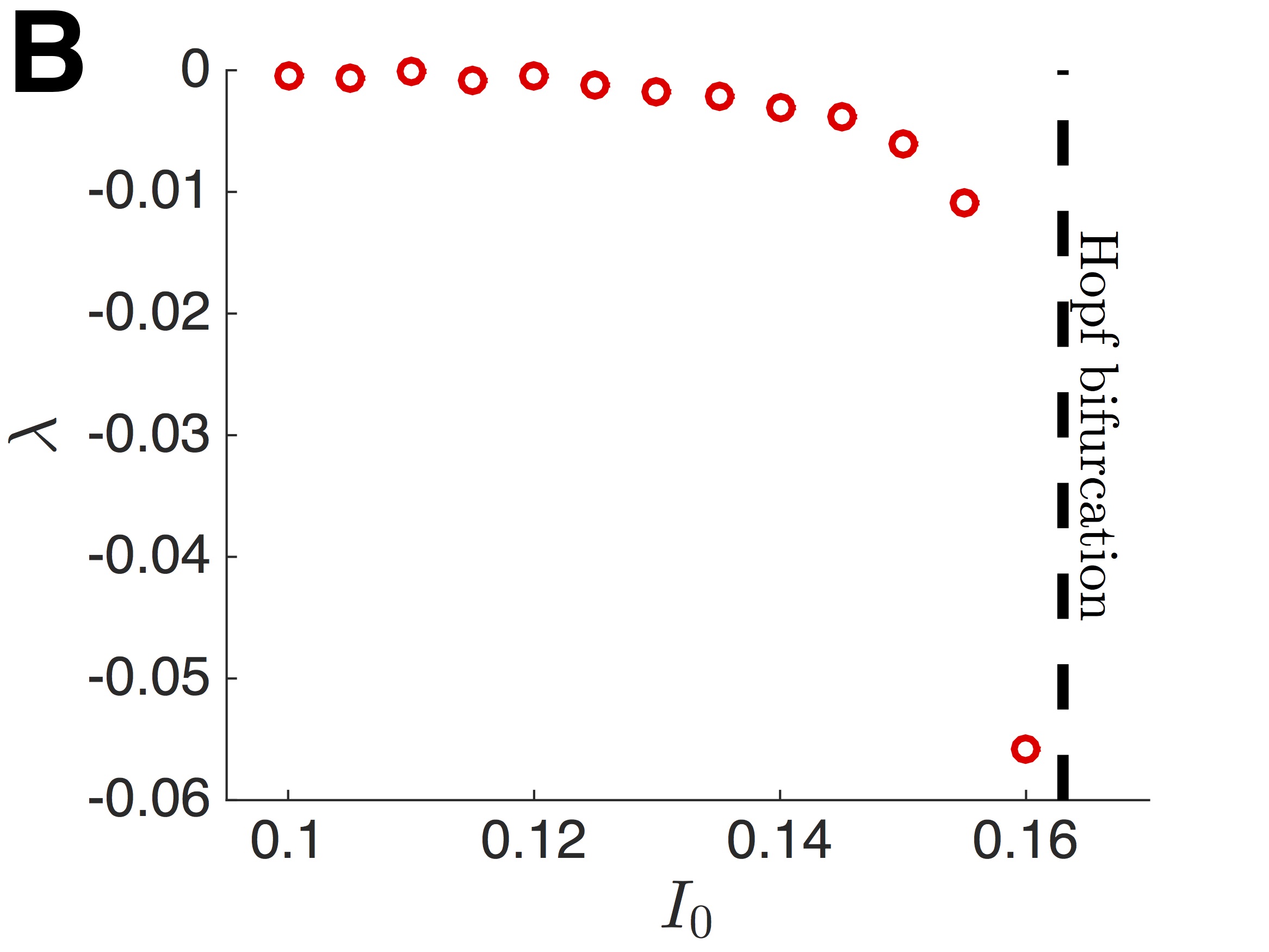} \end{center}
\vspace{-6mm}
\caption{(color online) Noise-driven synchronization of breather phases in a pair of adapting stochastic neural fields Eq.~(\ref{adaptsys}) with $w(x) = \cos (x)$ and $I(x)=I_0 \cos (x)$. ({\bf A}) Noise drives breather phases to the absorbing state where $\vartheta_1(t)=\vartheta_2(t)$ (see text); $I_0 = 0.1$. ({\bf B}) Numerically computed Lyapunov exponent $\lambda$ describing stability of phase-locked breathers increases in amplitude $|\lambda|$ with stimulus strength $I_0$. Parameters are $\alpha=0.1$, $\beta = 0.2$, $\ve = 0.04$. Other parameters are as in Fig. \ref{synchbump}.}
\vspace{-5mm}
\label{breathlock}
\end{figure}

\noindent
{\bf Breathers.} Lastly, we show that uncoupled oscillatory waves can also become phase-locked due to a common noise source. We extend Eq.~(\ref{layersys}) by incorporating linear adaptation as an auxiliary variable $q_j(x,t)$ ($j=1,2$) and a spatially varying external input $I(x)$ \cite{pinto01,folias05,ermentrout14}
\begin{subequations} \label{adaptsys}
\begin{align}
\d u_j &= \left[ -u_j - \beta q_j + w*f(u_j) + I \right] \d t + \ve \d W, \\
\dot{q}_j &= \alpha \cdot \left[ u_j - q_j \right],  \ \ \ \ \ j=1,2,
\end{align}
\end{subequations}
where $\alpha$ and $\beta$ are the rate and strength of adaptation. A detailed analysis of the onset of {\em breathers} in Eq.~(\ref{adaptsys}), via a Hopf bifurcation, can be found in \cite{folias05,ermentrout14}.

Our main interest is the rate at which breathers in a pair of uncoupled neural fields synchronize their phases when subject to common noise as in Eq.~(\ref{adaptsys}). An example of this phenomenon is shown in Fig. \ref{breathlock}{\bf A}. Note, we must take care in interpreting how the centers of mass of each oscillating bump relate to the phase of the underlying oscillation. In the case of bumps and waves, there was a one-to-one mapping between the wave positions $\Delta_j$ and the phase of the stochastically-driven oscillator. Here, we must track the activity $u_j$ and adaptation $q_j$ variables to resolve the phases $\vartheta_j$ of the underlying oscillations. Assuming breathers have period $T$, $u_j(x,t) = u_j(x,t+T)$ and $q_j(x,t) = q_j(x,t+T)$ so $\vartheta_j(t) = \vartheta_j (t+T)$, and there is a mapping $(u_j,q_j) \mapsto \vartheta_j$ for all values $(u_j,q_j)$ along the trajectory of a breather. We save a more detailed analytical determination of this mapping for future work. Here, we numerically determine $\vartheta_j(t)$, average, and compute the rate of decay $\lambda$ using the approximation $A + \lambda t \approx \frac{1}{N}\sum_{k=1}^N \ln | \vartheta_1(t) - \vartheta_2(t)|$ using least squares (Fig. \ref{breathlock}{\bf B}). Note, as in the case of bumps, the Lyapunov exponent increases in amplitude as it nears the pattern-generating (Hopf) bifurcation.    \\
\vspace{-2.5mm}

\noindent
{\bf Discussion.} Our results demonstrate common spatiotemporal fluctuations in neuronal networks can synchronize the phases of waves. We have shown this for stationary bumps, traveling waves, and breathers. Since our derivations mainly rely on our ability to derive an effective equation for the relative position of a noise-driven wave, we suspect we could extend them to the case of traveling fronts \cite{bressloff12b} or Turing patterns \cite{hutt07} in stochastic neural fields. Patterns on two-dimensional (2D) domains could also be addressed by deriving multidimensional effective equations for each pattern's position. For instance, a bump's position would be represented with a 2D vector \cite{poll14}, so there would be two Lyapunov exponents associated with bumps' phase-locked state. On the other hand, spiral waves would be characterized by a 2D position vector and a scalar phase \cite{laing05}, and phase/position-locked states would then have three Lyapunov exponents. We could also consider interlaminar coupling \cite{kilpatrick14}, exploring competing impacts of common noise and coupling on phase synchronization as in \cite{garcia09}. Furthermore, these results should be applicable to nonlinear PDE models of reaction-diffusion systems \cite{panja04,sagues07}. Overall, our results suggest a novel mechanism for generating coherent waves in laminar media, presenting a testable hypothesis that could be probed experimentally. \\
\vspace{-2.5mm}

\noindent
{\bf Appendix.} Here, we analyze the stochastic dynamics of bumps in a pair of uncoupled neural field models driven by {\em both common and independent noise sources}, extending previous work on limit cycle oscillators \cite{nakao07}. We incorporate an independent noise term into each layer of the stochastic neural field model \cite{kilpatrick14}:
\begin{align}
\d u_j(x,t) =& \left[ - u_j(x,t) + \int_{- \pi}^{\pi} w(x-y) f(u_j(y,t)) \d y \right] \d t \nonumber \\ & + \ve \left[ \chi \d W_c(x,t) + \sqrt{1 - \chi^2} \d W_j(x,t) \right].   \label{layersys2}
\end{align}
Small amplitude ($\ve \ll 1$) spatiotemporal noise terms $\d W_j (x,t)$ ($j=1,2,c$) are white in time and correlated in space, so $\langle \d W_j (x,t) \rangle = 0$; $\langle \d W_j(x,t) \d W_j(y,s) \rangle = 2 C_j(x-y) \delta (t-s) \d t \d s$ ($j=1,2,c$) with $C_j(x) = \sum_{k=0}^{\infty} a_k \cos (kx)$. The degree of correlation between layers is controlled by the parameter $\chi$.

Our analysis proceeds by considering stationary bumps in a network with even symmetric connectivity ($w(x) = w(-x)$). As in the main text, we characterize stochastic bump motion by applying the ansatz $u_j(x,t) = U(x - \Delta_j(t)) + \ve \Phi_j(x-\Delta_j(t),t) + {\mc O}(\ve^2)$, and $\Delta_j(0) = x_j$. Plugging this ansatz into Eq.~(\ref{layersys2}), expanding to ${\mc O}(\ve)$, and applying a solvability condition, we find that each $\Delta_j$ ($j=1,2$) obeys the Langevin equation
\begin{align*}
\d \Delta_j =& \ve \chi \frac{\int_{- \pi}^{\pi} V(x) \d W_c (x + \Delta_j,t) \d x}{\int_{- \pi}^{\pi}V(x) U'(x) \d x} \nonumber \\ & + \ve \sqrt{1 - \chi^2} \frac{\int_{- \pi}^{\pi} V(x) \d W_j (x + \Delta_j,t) \d x}{\int_{- \pi}^{\pi}V(x) U'(x) \d x}, 
\end{align*}
$j=1,2$, where the first and second term correspond to correlated and independent noise. Here, $V$ is a one-dimensional basis of ${\mc N}({\mc L}^*)$, where ${\mc L}^*p(x) = - p(x) + f'(U(x)) \int_{- \pi}^{\pi} w(x-y) p(y) \d y$. Since $U(x)$ is even symmetric, all components of the nullspace of ${\mc L}^*$ are necessarily odd symmetric \cite{kilpatrick13}. Note, we can represent $W_j(x,t) = a_0 X_0^{(j)} + \sum_{k=1}^{\infty} a_k \left[ X_k^{(j)} \cos (kx) + Y_k^{(j)} \sin (kx) \right]$ ($j=1,2,c$), where $X_k^{(j)}$ and $Y_k^{(j)}$ are normalized white noise processes. We can thus use trigonometric expansions to express
\begin{align}
\d \Delta_j = \sqrt{2} \ve \left[ \chi \d {\mc W}_c ( \Delta_j,t) + \sqrt{1- \chi^2} \d {\mc W}_j ( \Delta_j,t) \right],  \label{stratdel}
\end{align}
$j=1,2$, where ${\mc W}_j$ are multiplicative noise terms defined
\begin{align*}
{\mc W}_j(\Delta,t) = \sum_{k=1}^{\infty} \left[ b_{+k}^{(j)} \cos (k \Delta) X_k^{(j)} +  b_{-k}^{(j)} \sin (k \Delta) Y_k^{(j)} \right],
\end{align*}
and since $V(x)$ is odd symmetric, $X_0$ vanishes, and
\begin{align*}
b_{\pm k} = \pm \frac{a_k \int_{- \pi}^{\pi} V(x) \sin (kx) \d x}{\int_{- \pi}^{\pi} V(x) U'(x) \d x}.
\end{align*}
Eq.~(\ref{stratdel}) can be reformulated as an Ito equation $\d \Delta_j = B_j(\bd ) \d t + \d \zeta_j(\Delta_j,t)$, where $\zeta_j(\Delta_j,t) = \sqrt{2} \ve \left[ \chi {\mc W}_c ( \Delta_j,t) + \sqrt{1- \chi^2} {\mc W}_j ( \Delta_j,t) \right]$ ($j=1,2$) has correlations defined $\langle \d \zeta_j(\Delta_j,t) \d \zeta_k(\Delta_k,t) \rangle = {\mc C}_{jk} (\bd ) \d t$ ($j,k=1,2$), and the drift $B_j(\bd) = \frac{1}{4} \frac{\pd}{\pd \Delta_j} {\mc C}_{jj} (\bd)$ ($j=1,2$). Components of the correlation matrix are given
\begin{align*}
{\mc C}_{jk} ( \bd ) = 2 \ve^2 (\chi^2 +  \delta_{jk} \left( 1- \chi^2 \right)) \sum_{m=1}^{\infty} b_m^2 \cos [ m ( \Delta_j - \Delta_k)],
\end{align*}
where $\bd = (\Delta_1,\Delta_2)$ and $b_k = |b_{\pm k}|$, so it is straightforward to compute $B_j(\bd) \equiv 0$ ($j=1,2$).

The corresponding Fokker-Planck equation, describing the coevolution of the position variables $(\Delta_1,\Delta_2)$ is thus
\begin{align}
\frac{\pd P(\bd,t)}{\pd t} = & \  \ve^2 g(0) \left[ \frac{\pd^2 P(\bd,t)}{\pd \Delta_1^2} + \frac{\pd^2 P(\bd,t)}{\pd \Delta_2^2} \right] \label{fp1}  \\ & + 2 \ve^2 \chi^2 \frac{\pd^2}{\pd \Delta_1 \pd \Delta_2} \left[  g( \Delta_1 - \Delta_2) P(\bd,t) \right],  \nonumber
\end{align}
where $g( \Delta ) = \sum_{k=1}^{\infty} b_k^2 \cos ( k \Delta)$. Note, since $b_k^2 \geq 0, \forall k$, then $g(0) \geq g( \Delta)$ for $\Delta \in [ - \pi , \pi]$. We can write Eq.~(\ref{fp1}) as a separable equation by employing a change of variables that tracks the average $\rho = (\Delta_1 + \Delta_2)/2$ and phase difference $\phi = \Delta_1 - \Delta_2$ of the phase variables $\Delta_1$ and $\Delta_2$
\begin{align}
\frac{\pd P(\widetilde{\bd},t)}{\pd t} =& \ \ve^2 \left[ \frac{g(0)}{2} + \chi^2 g(\phi) \right] \frac{\pd^2 P(\widetilde{\bd},t)}{\pd \rho^2} \label{fp2}  \\ 
& + 2 \ve^2 \frac{\pd^2}{\pd \phi^2} \left( \left[g(0) - \chi^2 g(\phi) \right] P(\widetilde{\bd},t) \right), \nonumber
\end{align}
where $\widetilde{\bd} = (\rho, \phi)$. Eq.~(\ref{fp2}) can be decoupled by plugging in the ansatz $P(\widetilde{\bd},t) = S(\rho,t) \cdot M(\phi,t)$ and noting the equation will be satisfied by the system
\begin{align}
\frac{\pd S(\rho,t)}{\pd t} &= \ve^2 \left[ \frac{g(0)}{2} + \chi^2 g(\phi) \right] \frac{\pd^2 S(\rho,t)}{\pd \rho^2}, \label{fpsys} \\
\frac{\pd M(\rho,t)}{\pd t} &= 2 \ve^2 \frac{\pd^2}{\pd \phi^2} \left( \left[ g(0) - \chi^2 g(\phi) \right] M(\phi,t) \right). \nonumber
\end{align}
Thus, we can solve for the stationary solution of the system, Eq.~(\ref{fpsys}), by setting $S_t = M_t \equiv 0$ and requiring periodic boundary conditions. The stationary distribution for the position average is $S_0(\rho) = 1/(2 \pi)$. Furthermore, we can integrate the stationary equation for $M(\phi,t)$ to find that the stationary density of the phase difference is
\begin{align}
M_0(\phi) = \frac{m_0}{g(0) - \chi^2 g(\phi)},
\end{align}
where $m_0 = 1/ \int_{- \pi}^{\pi} \left[ g(0) - \chi^2 g(x) \right]^{-1} \d x$ is a normalization factor. When noise to each layer is independent ($\chi \to 0$, uncorrelated), then $M_0(\phi) = 1/ 2\pi$ is constant in space. Since no common noise source entrains the phase of each bump, the bumps diffuse independently of one another. However, when noise is totally correlated between layers ($\chi \to 1$), then $M_0(\phi) = \delta (\phi)$. Thus, all initial conditions eventually result in the phase-locked state $\Delta_1 = \Delta_2$. The stationary distribution $M_0(\phi)$ broadens as $\chi$ is decreased, with a peak still remaining at $\phi = 0$.

To compare our results to numerical simulations, we compute the stationary density $M_0(\phi)$ explicitly by using $f(u) = H(u - \theta)$; $w(x) = \cos (x)$; and $C_j(x) = \cos (x)$ ($j=1,2,c$). Stable stationary bumps $U(x) = 2 \sin a \cos (x)$ satisfy the threshold condition $U(\pm a) = \theta$, with half-width $a$. We can thus compute the null vector $V(x) = \delta (x-a) - \delta (x+a)$ and find $b_{\pm 1} = \mp 1/ \left[ \sqrt{1+ \theta} + \sqrt{1 - \theta} \right]$ and $b_{\pm k} \equiv 0$, $k \neq 1$. Therefore
\begin{align}
M_0 (\phi) = \frac{\sqrt{1 - \chi^4}}{2 \pi \left[ 1 - \chi^2 \cos (\phi) \right]}.
\end{align}
\vspace{-2.5mm}

ZPK was funded by NSF-DMS-1311755. We thank Oliver Langhorne for helpful conversations.

\bibliographystyle{apsrev}
\bibliography{synchwave}

\begin{thebibliography}{36}
\bibitem[{num()}]{numerics}
\bibinfo{note}{Numerical simulations used Euler-Maruyama method along with a
  direct spectral method for evaluating integrals. Timesteps $dt = 0.01$ and
  spatial steps $dx = 2 \pi / 1000$. Numerically evaluated Lyapunov exponents
  were determined by averaging $\ln | \Delta_1(t) - \Delta_2(t)|$ across
  $N=5000$ realizations and fitting to $A + \lambda \cdot t$ using MATLAB's
  polyfit function.}
\end{thebibliography}

\end{document}